\documentstyle[prl,aps,preprint]{revtex}

\def\cqg{Class.\ Quant.\ Grav.\ }

\def\plb{Phys.\ Lett.\ B}
\def\ii{\'{\i}}
\def\bi{\bigskip}
\def\noi{\noindent}
\def\be{\begin{equation}}
\def\en{\end{equation}}
\def\bq{\begin{eqnarray}}
\def\eq{\end{eqnarray}}
\begin{document}

\begin{center}
{\Large \bf Supersymmetric Action for FRW Model with Complex Matter
Field}\\[1.5cm]
\end{center}
\vspace{.7cm}
\begin{center}
{{\large \bf V.I. Tkach$^{1}$\footnote{e-mail vladimir@ifug1.ugto.mx}
and J.J. Rosales$^{1,2}$\footnote{e-mail: juan@ifug3.ugto.mx}}\\[1cm]
\vspace{.5cm}
{\it $^{1}$ Instituto de F\'\i sica, Universidad de Guanajuato}\\
{\it Apartado Postal E-143, Le\'on, Gto., M\'exico} \\
{\it and} \\
{\it $^{2}$ Universidad Aut\'onoma Metropolitana-Iztapalapa}\\
{\it Depto. de F\ii sica, Apdo. Postal 55-534} \\
{\it M\'exico, D.F. M\'exico}}
\end{center}

\vspace{1.5cm}

\begin{center}
{\bf Abstract}
\end{center}

On the basis of the local $n=2$ supersymmetry we construct the supersymmetric
action for a set of complex scalar supermultiplets in the FRW model. This
action corresponds to the dilaton-axion and chiral components of supergravity
theory.

\bi

PACS Number(s): 04.60. Kz, 04.65. + e, 12.60. Jv., 98.80. Hw.

\bi
\bi

\setlength{\baselineskip}{1\baselineskip}

\newpage

One of the possible models of unification of interactions is supergravity
interacting with matter, because it admits the solutions of the problem of
cosmological constant and comes to degeneration of masses for fields in any
supermultiplet [1].

In the last years spatially homogeneous minisuperspaces
models have to be indeed a
very valuable tool in supergravity theories. The study of minisuperspace have
led to important and interesting results, pointing out usefull lines of
research. Since then, several publications have appeared on the subject of
supersymmetric quantum cosmology, also including matter [2].

More recently we proposed a new formulation to investigate supersymmetric
quantum cosmological models [3,4]. This formulation was performed by
introducing a superfield formulation. This is because superfields defined on
superspace allow all the component fields in a supermultiplet to be
manipulated simultaneously in a manner, which automatically preserves
supersymmetry. Our approach has the advantage of being more simple, than
proposed models based on full supergravity [2] and gives, by means of this
local symmetry procedure, in a direct maner the  corresponding fermionic
partners.

In the paper [5] was considered the FRW model interacting with a
simplest real matter supermultiplet, and was shown, that in this model of
Universe with local supersymmetry, when energy density of vacuum is equal to
zero, allows to have breaking supersymmetry.

In this work we will
consider the homogeneous and isotropic FRW cosmological model interacting with
a set of homogeneous and isotropic complex scalar matter supermultiplet. As
will be shown, the supersymmetric action obtained for this model corresponds
to dilaton-axion and chiral components of supergravity theory.
It is well known, that the action of cosmological FRW models is
invariant under
the time reparametrization $t^\prime \to t+a(t)$. Then, we may obtain the
superfield description of these models, when we introduce the odd complex
``time" parameters $\eta ,\bar\eta$, which are the superpartners of time
parameter [3,4]. This procedure is well known from superparticles [6-8].

So, we have the following superfield action for the FRW model
interacting with
a set of complex scalar matter supermultiplet
\bq
&& S=\int\bigg\{-\frac{{I\!\!N^{-1}}}{2\kappa^2} {I\!\!R}\bar D_\eta {I\!\!R}
D_\eta {I\!\!R}+\frac{\sqrt{k}}{2\kappa^2} {I\!\!R}^2 +\nonumber \\
&+& \frac{{I\!\!N}^{-1} {I\!\!R}^3}{4}\bigg(\bar D_\eta \bar Z^a D_\eta Z^a+
\bar D_\eta Z^a D_\eta \bar Z^a\bigg) -{I\!\!R}^3 |g(Z)|\bigg\}
d\eta d\bar\eta dt ,
\eq
\noi where $\kappa=\sqrt{\frac{4\pi G}{3}}$ and $G$ is the Newtonian constant
of gravity. In the action (1) ${I\!\!N}(t,\eta ,\bar\eta)$ is a real gravity
superfield, and it has the form
\be
{I\!\!N} (t,\eta ,\bar\eta) = N(t) +i\eta \psi^\prime(t)+i\bar\eta \bar
\psi^\prime  (t) + \eta \bar\eta V^\prime (t),
\en
\noi where $\psi^\prime (t) = N^{1/2} \psi (t), \bar\psi^\prime (t) = N^{1/2}
\bar\psi (t)$ and $V^\prime (t) = NV -\bar\psi \psi$. The law transformation
of the superfield ${I\!\!N}(t,\eta ,\bar\eta)$ may be written in the following
way
\be
\delta{I\!\!N} (t,\eta ,\bar\eta)=({I\!\!\Lambda}{I\!\!N})^.
+ \frac{i}{2} \bar D_\eta {I\!\!\Lambda} D_\eta {I\!\!N} + \frac{i}{2} D_\eta
{I\!\!\Lambda} \bar D_\eta {I\!\!N} ,
\en
\noi where the superfunction ${I\!\!\Lambda}(t,\eta ,\bar\eta)$ is written as
\be
{I\!\!\Lambda} (t,\eta,\bar\eta)=a(t)+i\eta\beta^\prime (t)
+i\bar\eta\bar\beta^\prime (t) +\eta\bar \eta b (t) ,
\en

\noi and $D_\eta = \frac{\partial}{\partial\eta}
+ i \bar \eta \frac{\partial}{\partial t}$ and $\bar D_\eta =
- \frac{\partial}{\partial\bar\eta} - i \eta
\frac{\partial}{\partial t}$ are the supercovariant derivatives, $\beta^\prime
(t) = N^{-1/2} \beta (t)$ is the Grassmann complex parameter of the local
``small" $n=2$ susy transformations and $b (t)$ is the parameter of local
$U(1)$ rotations of the complex $\eta$.

\noi So, the $n=2$ local transformations of the supertime $(t,\eta ,\bar\eta)$
are
\bq
\delta t &=& {I\!\!\Lambda}(t,\eta ,\bar\eta)+\frac{1}{2}\bar\eta\bar D_\eta
{I\!\!\Lambda} (t, \eta , \bar \eta) -\frac{1}{2} \eta D_\eta {I\!\!\Lambda}
(t, \eta ,\bar\eta), \nonumber \\
\delta \eta &=& \frac{i}{2} \bar D_\eta {I\!\!\Lambda} (t,\eta , \bar\eta)
, \\
\delta\bar\eta &=&-\frac{i}{2}D_\eta{I\!\!\Lambda}(t,\eta ,\bar\eta),\nonumber
\eq

\noi which are the generalization of the time reparametrization $t^\prime \to
t+a (t)$ in the cosmological models.

\noi The components of the superfield ${I\!\!N} (t,\eta ,\bar\eta)$ in (2) are
gauge fields of the one-dimensional $n=2$ extended supergravity, $N(t)$ is
einbein, $\psi (t)$ and $\bar\psi (t)$ are the time-like components of the
Rarita-Schwinger fields $\psi^\alpha_\mu$ and $\bar\psi^\alpha_\mu$, which may
be obtained by spatial reduction from the four dimensional supergravity to the
one-dimensional models and $V (t)$, is $U(1)$ gauge field.

The real ``matter" superfield ${I\!\!R} (t,\eta ,\bar\eta)$ may be written as
\be
{I\!\!R} (t,\eta ,\bar\eta)=R (t)+i\eta \lambda^\prime (t)+i\bar\eta
\bar\lambda^\prime (t) +\eta \bar\eta B^\prime (t) ,
\en
\noi where $\lambda^\prime (t)=\kappa N^{1/2} \lambda (t), \bar\lambda^\prime
(t) =\kappa N^{1/2} \bar\lambda (t)$ and $B^\prime (t) = \kappa NB -
\frac{\kappa}{2} (\bar\psi \lambda - \psi \bar\lambda ).$
The law tranformation for the superfield ${I\!\!R} (t,\eta ,\bar\eta)$ is
\be
\delta {I\!\!R} = {I\!\!\Lambda} \dot{I\!\!R}+\frac{i}{2} \bar D_\eta
{I\!\!\Lambda} D_\eta {I\!\!R} +\frac{i}{2} D_\eta{I\!\!\Lambda}\bar D_\eta
{I\!\!R} .
\en
\noi The  component $B(t)$ in (6) is an auxiliary degree of freedom, $\lambda
(t)$ and $\bar\lambda (t)$ are dynamical degrees of freedom remaining from
the
spatial part of the Rarita-Schwinger field obtained from spatial reduction of
pure supergravity theories to cosmological models and are the partners of the
scale factor $R(t)$.

\noi The complex scalar matter supermultiplet $Z^a$ consists
of a set of spatially homogeneous scalar complex matter fields $z^a(t)$,
$\bar z^a(t) ~(a=1,\ldots n)$, four fermionic degrees of freedom $\chi^a(t)$,
$\bar \chi^a (t)$, $\phi^a (t)$ and $\bar\phi^a (t)$, two bosonic auxiliary
fields $F^a(t)$ and $\bar F^a (t)$ and a superpotential of the type $|g(Z)|$.
The components of the complex matter superfields may be written in the
following way
\be
Z^a (t,\eta ,\bar\eta)= z^a (t) +i\eta \chi^{\prime a} (t) +i\bar\eta \bar
\phi^{\prime a} (t) + F^{\prime a} (t) \eta\bar\eta ,
\en
\noi where $\chi^{\prime a}{(t)} = N^{1/2} \chi^a (t), \bar \phi^{\prime a}
(t) = N^{1/2} \bar \phi^a (t)$ and $F^{\prime a} (t) = N F^a- \frac{1}{2}
(\bar\psi^a \chi^a - \psi^a \bar \chi^a)$. The law transformation for the
complex matter superfield is written as
\be
\delta Z^a= {I\!\!\Lambda} \dot Z^a +\frac{i}{2} \bar D_\eta {I\!\!\Lambda}
D_\eta Z^a +\frac{i}{2} D_\eta {I\!\!\Lambda} \bar D_\eta Z^a .
\en

\noi It is clear, that the superfield action (1) is invariant under the $n=2$
local super time transformations (5) if the superfields transform as (3,7,9).

We will write the expression, which is found under the integral (1) by
means of certain superfunction $f({I\!\!R}, {I\!\!N}, Z^a)$. Then, the
infinitesimal small transformation of the action (1) under the superfield
transformations (3,7,9) has the following form:
\be
\delta S=\frac{i}{2} \int\bigg\{ \bar D_\eta \bigg( {I\!\!\Lambda} D_\eta
f({I\!\!R}, {I\!\!N}, Z^a)\bigg) + D_\eta \bigg( {I\!\!\Lambda}\bar D_\eta
f({I\!\!R}, {I\!\!N}, Z^a) \bigg) \bigg\} d\eta d\bar\eta dt .
\en

We can see, that under the integration it gives a total derivative. That
is, the action (10) is invariant under the superfield transformations
(3,7,9).

\noi Making the corresponding operations from the action (1) one obtains the
expression for the component action, where the auxiliary fields $B (t),
F^a(t)$
and $\bar F^a(t)$ appear. Performing the variation with respect to these
auxiliary fields we get three algebraical equations, which have the
following solutions
\be
B(t)= \frac{\kappa}{2R} \bar\lambda\lambda + \frac{\sqrt{k}}{\kappa} +
\frac{3\kappa}{4} R (\bar \chi^a \chi^a +\bar\phi^a \phi^a)-3\kappa R|g| ,
\en
\be
F^a(t) =- \frac{3\kappa}{2R} (\lambda \bar\phi^a -\bar\lambda \chi^a)
+ 2 \frac{\partial |g|}{\partial \bar z_a} ,
\en
\noi and
\be
\bar F^a (t) =- \frac{3\kappa}{2R} (\lambda \bar \chi^a -\bar\lambda \phi^a)+2
\frac{\partial |g|}{\partial z_a} ,
\en

\noi and after substituting them again into the component action obtained
from (1) and making the following fields reedefinitions $\lambda\to R^{-1/2}
\lambda$, $\bar\lambda \to R^{-1/2} \bar\lambda$, $\chi^a \to 2^{1/2} R^{-3/2}
\chi^a$, ~$\bar \chi^a ~\to ~2^{1/2} ~R^{-3/2} ~\bar \chi^a$, ~$\phi^a ~\to
~2^{1/2} ~R^{-3/2}~\phi^a$ ~and ~$\bar\phi^a ~\to ~2^{1/2} ~R^{-3/2}
~\bar\phi^a$. We get the following component action

\bq
S &=& \int \bigg\{ -\frac{R}{2N\kappa^2}(DR)^2+ \frac{R^3}{2N} D\bar z^a
 D z^a
+\frac{NRk}{2\kappa^2}+\frac{9}{2}N\kappa^2 R^3|g(z)|^2 -\nonumber \\
&-&\frac{NR^3}{2}\frac{\partial g}{\partial z^a}\frac{\partial g}{\partial\bar
z^a}-3N\sqrt{k} R^2|g|+i\bar\lambda D\lambda -i\bar\phi^a D\phi^a -
i\bar\chi^a D\chi^a \nonumber \\
&+& \frac{3\sqrt{2}\kappa}{4}iD z^a (\lambda \bar\chi^a +
\bar\lambda \phi^a)+ \frac{3\sqrt{2}\kappa}{4}i D \bar z^a (\bar\lambda \chi^a
+ \lambda \bar\phi^a) -\nonumber \\
&-& \frac{\sqrt{k}\sqrt{R}}{2\kappa} (\bar\psi\lambda -\psi\bar\lambda)
- \frac{N\sqrt{k}}{2R} \bar\lambda \lambda + \frac{3N\sqrt{k}}{2R}
(\bar\phi^a \phi^a
+ \bar\chi^a \chi^a) + \frac{9N\kappa^2}{4R^3} \bar\chi^a \chi^a \bar\phi^a
\phi^a  \nonumber \\
&-& \frac{3\sqrt{2}}{4R^{3/2}} \kappa (\bar\psi \lambda
- \psi\bar\lambda)(\bar\phi^a \phi^a +\bar\chi^a \chi^a)
- \frac{9}{2} N \kappa^2 |g|(\bar\chi^a \chi^a
+ \bar\phi^a \phi^a) \\
&+& 2N \frac{\partial^2 |g|}{\partial z_a \partial z_b}
\bar\phi^a \chi^b + 2N\frac{\partial^2|g|}{\partial\bar z_a\partial \bar z_b}
 \bar\chi^a \phi^b +\frac{9}{2}\kappa^2N|g|\bar\lambda\lambda \nonumber \\
&+& \frac{3\sqrt{2}}{2}\kappa N\bigg[\frac{\partial|g|}{\partial z_a}
(\bar\phi^a\lambda +\bar\lambda\chi^a)+ \frac{\partial |g|}{\partial\bar z_a}
(\bar\chi^a\lambda +\bar\lambda \phi^a) \bigg] + 2N\frac{\partial^2|g|}
{\partial z_a\partial \bar z_b}(\bar\phi^a \phi^b + \bar\chi^a \chi^b)
\nonumber \\
&+& \frac{\bar\psi}{2} \bigg[ \sqrt{2} R^{3/2} \frac{\partial
|g|}{\partial z_a} \chi^a + \sqrt{2} R^{3/2} \frac{\partial |g|}{\partial \bar
z_a} \phi^a + 3\kappa R^{3/2} |g| \lambda \bigg]\nonumber\\
&-& \frac{\psi}{2} \bigg[ \sqrt{2} R^{3/2}
\frac{\partial |g|}{\partial z_a} \bar\phi^a +\sqrt{2} R^{3/2} \frac{\partial
|g|}{\partial \bar z_a} \bar\chi^a+3\kappa R^{3/2} |g| \bar\lambda \bigg]
\bigg\} dt , \nonumber
\eq

\noi where $DR=\dot R - \frac{i\kappa}{2\sqrt{R}} (\bar \psi \lambda + \psi
\bar\lambda) , D z^a = \dot z^a -\frac{i}{\sqrt{2}\sqrt{R^3}}(\psi \bar\phi^a
+
\bar\psi\chi^a)$ and $D\bar z^a={\dot{\bar{z}}}^a-\frac{i}{\sqrt{2}\sqrt{R^3}}
(\bar\psi \phi^a +\psi \bar\chi^a)$ are the supercovariant derivatives,
$D\lambda = \dot\lambda -\frac{i}{2} V\lambda$ and $D\phi^a = \dot\phi^a +
\frac{i}{2} V\phi^a$ are the covariant derivatives.

>From (1) we can see, that the action for a set of scalar complex supermatter
has the form

\be
\int \bigg\{ \frac{1}{4} \bigg( \bar D_\eta \bar Z^a D_\eta Z^a + \bar D_\eta
Z^a D_\eta \bar Z^a\bigg) - |g (Z)|\bigg\} d\eta d\bar\eta dt .
\en

\noi This action corresponds to the action obtained by spatial reduction from
Wess-Zumino model in four dimensions with arbitrary superpotential $g(Z)$. The
action (15) gives two complex supercharges $Q_1$ and $Q_2$. Because of
the action (15) is invariant under the change of $Z^a\leftrightarrow\bar Z^a$,
then the supercharges allow the invariance under the change of
$Q_1 \leftrightarrow\bar Q_2$. Furthermore, we can join in a one complex
supercharge $\widetilde{S}=Q_1 + \bar Q_2$ and $\bar{\widetilde{S}}=\bar
{Q}_1 + Q_2$.

Now we will proceed with the hamiltonian analysis of the system. The momenta
$\Pi_R, \Pi^a_z$ and $\Pi^a_{\bar z}$ conjugate to $R(t), z^a (t)$ and
$\bar z^a(t)$ respectively, they are given by

\bq
\Pi_R &=&- \frac{R}{N\kappa^2} \bigg[ \dot R -\frac{i}{2\sqrt{R}}
(\psi \bar\lambda + \bar\psi \lambda )\bigg] , \\
\Pi^a_z &=& \frac{R^3}{N} \bigg[ {\dot{\bar{z}}}^a-\frac{i}{\sqrt{2}R^{3/2}}
(\psi \bar\chi^a +\bar\psi \phi^a)\bigg] + \frac{3i\sqrt{2}\kappa}{4R^{3/2}}
(\lambda \bar\chi^a +\bar\lambda \phi^a) , \\
\Pi^a_{\bar z} &=& \frac{R^3}{N} \bigg[ \dot z^a - \frac{i}{\sqrt{2}R^{3/2}}
(\bar\psi \chi^a +\psi\bar\phi^a)\bigg] +\frac{3i\sqrt{2}\kappa}{4R^{3/2}}
(\bar\lambda \chi^a + \lambda \bar\phi^a )
\eq

\noi with respect to the canonical Poisson brackets

\be
\{R, \Pi_R \} =1,\{ z_a ,\Pi^b_z\}=\delta^b_a , \{\bar z_a ,\Pi^b_{\bar z}\}
= \delta^b_a .
\en

\noi As usual with fermionic systems the calculation of the momenta conjugate
to the anticommuting dynamical variables introduces primary constraints, which
we denote by
\bq
\sqcap_\lambda \equiv \Pi_\lambda +\frac{i}{2} \bar\lambda \approx 0, &~&
\sqcap_{\bar\lambda} \equiv \Pi_{\bar\lambda}+\frac{i}{2} \lambda \approx 0
,\nonumber \\
\sqcap^a_\chi \equiv \Pi_\chi^a -\frac{i}{2} \bar\chi^a \approx 0, &~&
\sqcap^a_{\bar\chi} \equiv \Pi_{\bar\chi}^a -\frac{i}{2} \chi^a \approx 0 ,\\
\sqcap^a_\phi \equiv \Pi^a_\phi -\frac{i}{2} \bar\phi^a \approx 0, &~&
\sqcap^a_{\bar\phi} \equiv \Pi^a_{\bar\phi}- \frac{i}{2} \phi^a \approx 0 ,
\nonumber
\eq

\noi where $\Pi_\lambda =\frac{\partial L}{\partial \dot\lambda} ,\Pi^a_\chi =
\frac{\partial L}{\partial\dot\chi_a}$ and $\Pi^a_\phi =
\frac{\partial L}{\partial \dot\phi_a}$ are the momenta conjugate to the
anticommuting dynamical
variables $\lambda (t)$, $\chi (t)$ and $\phi (t)$ respectively. The momenta
conjugate to $B(t), \psi (t), \bar\psi (t)$ and $V (t)$ are found equal to
zero indicating, that these variables play the role of gauge fields, whose
time derivative is arbitrary, so they are non-dynamical fields.

The constraints (20) are of second-class and can be eliminated by Dirac
procedure. We define the matrix constraint $C_{ik} (i,k=\lambda ,\bar\lambda ,
\chi^a,\bar\chi^a ,\phi^a, \bar\phi^a)$ as the Poisson bracket. We have the
following non-zero matrix elements

$$C_{\bar\lambda\lambda} = C_{\lambda\bar\lambda}= \{ \sqcap_{\lambda a},
\sqcap_{\bar\lambda}\,^b \} = i\delta^b_a \quad ,
\quad C_{\bar\chi\chi} = C_{\chi\bar\chi} = \{ \sqcap_{\chi a} ,
\sqcap_{\bar\chi}^b \} =i\delta^b_a  ,$$
\be
C_{\bar\phi\phi} = C_{\phi\bar\phi} = \{ \sqcap_{\phi a},
\sqcap_{\bar\phi}\,^b\} = i \delta^b_a
\en

\noi with their inverse matrices $(C^{-1})^{\bar\lambda\lambda} =-i,
(C^{-1})^{\bar\chi\chi} =i$ and $(C^{-1})^{\bar\phi\phi} =i$. The Dirac
brackets $\{~,\}^*$ are then defined by

\be
\{ A,B\}^* = \{A,B\} -\{A, \sqcap_i\} (C^{-1})^{ik} \{\sqcap_{ki} B \} .
\en

The result of this procedure is the elimination of the momenta
conjugate to the
fermionic variables, leaving as with the following non-zero Dirac brackets
relations

$$\{ R, \Pi_R\}^* =\{R, \Pi_R\}=1 \quad , \quad \{ z_a \Pi^b_z\}^* =\{ z_a,
\Pi^b_z \} =\delta^b_a ,$$
\be
\{ \bar z_a. \Pi^b_{\bar z} \}^* = \{ \bar z_a , \Pi^b_{\bar z}\}=\delta^b_a ,
\en
$$\{\lambda ,\bar\lambda\}^* =i \quad , \quad \{ \chi_a, \bar\chi^b \}^*
=-i\delta^b_a \quad , \quad \{\phi_a , \bar\phi^b \}^* =-i\delta^b_a .$$

In a quantum theory Dirac brackets $\{~~,~~\}^*$ must be replaced by
commutators [~,~] or anticommutators $\{~~,~~\}$. We get

$$[R,\Pi_R]=i\{R,\Pi_R\}^*=i \quad ,\quad  [z_a,\Pi^b_z]=i\{z_a, \Pi^b_z\}^* =
i\delta^b_a,$$
\be
[\bar z_a, \Pi^b_{\bar z}] =i \{\bar z_a, \Pi^b_{\bar z}\}^* =i\delta^b_a
\en
$$\{\lambda, \bar\lambda \}=i \{\lambda ,\bar\lambda\}^* =-1 \quad , \quad
\{\chi_a , \bar\chi^b\}=i [\chi_a, \bar\chi^b\}^* = \delta^b_a ,$$
$$ \{\phi_a ,\bar\phi^b\} =i \{ \phi_a \bar\phi^b \}^* =\delta^b_a ,$$
\noi where we choose the unity $\hbar = c = 1$. The first class constraints
may be obtained from the action (14), varging $N(t), \psi(t),
\bar\psi(t)$ and
$V(t)$ respectively. We obtain the following first-class constraints.
\bq
H &=& -\frac{\kappa^2 \Pi^2_R}{2R} -\frac{kR}{2\kappa^2}
+\frac{2}{R^3}\Pi^a_z \Pi^a_{\bar z}-\frac{9\kappa^2}{2} R^3 |g(z)|^2 +
\frac{R^3}{2} \frac{\partial g}{\partial z^a} \frac{\partial \bar g}{\partial
 \bar z_a} +\frac{\sqrt{k}}{2R} \bar\lambda\lambda  \nonumber\\
&-&\frac{3\sqrt{2}}{2}i \kappa\frac{\Pi^a_z}{R^3}(\lambda\bar\phi^a
+\bar\lambda \chi^a)-  \frac{3\sqrt{2}}{2} \frac{i\kappa \Pi^a_{\bar z}}{R^3}
 (\bar\lambda \phi^a
+\lambda\bar\chi^a)-\frac{9\kappa^2}{4R^4}(\bar\phi^a\phi^a
+\bar\chi^a \chi^a)
\bar\lambda\lambda \nonumber\\
&-& \frac{3\sqrt{k}}{2R} ( \bar \phi^a \phi^a + \bar \chi^a \chi^a)
+\frac{9\kappa^2}{2} |g|\bar\lambda\lambda -\frac{9\kappa^2}{4R^3} \bar\phi^a
\phi^a \bar\chi^a \chi^a + \frac{9\kappa^2}{2} |g|(\bar\phi^a\phi^a
+\bar\chi^a
\chi^a) \\
&-& 2 \frac{\partial^2|g|}{\partial z_a\partial z_b}\bar\phi^a \chi^b -2
\frac{\partial^2 |g|}{\partial \bar z_a \partial \bar z_b} \bar \chi^a \phi^b
- \frac{3\sqrt{2}}{2} \kappa \frac{\partial |g|}{\partial\bar z_a} (\bar
\lambda \phi^a + \bar \chi^a \lambda )\nonumber\\
&-& \frac{3\sqrt{2}}{2} \kappa \frac{\partial |g|}{\partial z_a}
(\bar\lambda \chi^a +\bar\phi^a \lambda) -2
\frac{\partial^2 |g|}{\partial z_a \partial \bar z_b} (\bar\phi^a \phi^b +
\bar\chi^a \chi^b) ,\nonumber
\eq
\bq
S &=& \bigg[\frac{\kappa \Pi_R}{\sqrt{R}} - \frac{i\sqrt{k}\sqrt{R}}{\kappa} -
\frac{3i\kappa}{2\sqrt{R^3}} (\bar\phi^a \phi^a + \bar\chi^a \chi^a)+3i\kappa
\sqrt{R^3} |g| \bigg] \lambda + \\
&+& \bigg[ \frac{\sqrt{2}}{\sqrt{R^3}} \Pi^a_z + i \sqrt{2} \sqrt{R^3}
\frac{\partial |g|}{\partial z^a}\bigg]\chi^a+\bigg[\frac{\sqrt{2}}
{\sqrt{R^3}}
\Pi^a_{\bar z} +i \sqrt{2} \sqrt{R^3} \frac{\partial |g|}{\partial \bar z_a}
\bigg] \phi^a , \nonumber \\
\bar S&=& \bigg[\frac{\kappa \Pi_R}{\sqrt{R}}+\frac{i\sqrt{k}\sqrt{R}}{\kappa}
+\frac{3i\kappa}{2\sqrt{R^3}} (\bar\phi^a \phi^a + \bar\chi^a \chi^a )- 3 i
\kappa \sqrt{R^3} |g| \bigg] \bar\lambda + \\
&+& \bigg[ \frac{\sqrt{2}}{\sqrt{R^3}} \Pi^a_{\bar z} -i \sqrt{2}\sqrt{R^3}
\frac{\partial |g|}{\partial \bar z_a} \bigg] \bar \chi^a + \bigg[ \sqrt{2}
\frac{\Pi^a_z}{\sqrt{R^3}} -i \sqrt{2} \sqrt{R^3} \frac{\partial |g|}{\partial
z_a} \bigg] \bar\phi^a , \nonumber
\eq
\noi and
\be
{\cal F} = (-\bar\lambda \lambda + \bar\phi^a \phi^a +\bar\chi^a \chi^a) .
\en

The constraints (25,28) follow from invariant action (14) under the ``small"
local supertransformations (5). The general hamiltonian is a sum of all the
constraints, {\it i.e.}

\be
H_G=NH+\frac{i}{2}\psi\bar S+\frac{i}{2}\bar\psi S+{\cal F}(\frac{1}{2}\vee ).
\en

In the quantum theory the first class constraints associated with the
invariance of the action (14) become conditions on the wave function with the
commutation rule (24), so that any physically allowed state must obey the
following quantum constraints.
\be
H|\psi>=0 , \quad S|\psi>0 , \quad \bar S|\psi>=0, \quad {\cal F}|\psi>=0 ,
\en

\noi which are obtained when we change the classical dynamical variables by
operators $\Pi_R =-i \frac{\partial}{\partial R} , \Pi^a_Z =-i
\frac{\partial}{\partial Z_a}$ and making the following matrix
representation for fermionics
variables $\lambda , \bar\lambda ,\chi , \bar\chi , \phi$ and $\phi$,
\bq
\lambda =-\sigma^-\otimes 1\otimes 1 , &~& \bar\lambda =\sigma^+ \otimes 1
\otimes 1 , \nonumber \\
\chi = \sigma^3 \otimes \sigma^- \otimes 1 , &~& \bar\chi = \sigma^3 \otimes
\sigma^+ \otimes 1 , \\
\phi = \sigma^3 \otimes \sigma^3 \otimes \sigma^+ , &~& \bar\phi =
\sigma^3 \otimes \sigma^3 \otimes \sigma^- , \nonumber
\eq
\noi where $\sigma^\pm = \frac{\sigma_1 \pm i \sigma_2}{2}$ and $\sigma_1 ,
\sigma_2$ and $\sigma_3$ are the Pauli matrices. For the quantum generators
$H, S, \bar S$ and ${\cal F}$ we obtain the following closed superalgebra

\be
\{S,\bar S\} =2H,\qquad\qquad [S,H]=0,\qquad\qquad [{\cal F}, S]=-S,
\en
$$S^2 =\bar S^2 = 0,\qquad\qquad [\bar S, H]=0,\qquad\qquad [{\cal F}, \bar S]
= \bar S,$$

\noi where $H$ is the hamiltonian of the system, $S$ is the single complex
supersymmetric charge of the $n=2$ supersymmetric quantum mechanics and
${\cal F}$ is the fermion number operator.

\bi
\bi

\noi {\bf Conclusion}

\bi

Hence, on the base of the local ``small" supersymmetry we considered the FRW
cosmological model with the set of complex superfields and a superpotential
$g(Z^a)$. As effective supergravity theory coupled to ``observable" sector
with gauge group $SU(3) \times SU(2) \times U(1)$ through a ``hidden" sector
[9] corresponding to four-dimensional superstrings, the next step will be
inclusion of K\"ahler function $K(Z^a,\bar Z^b)$ to the scheme for chiral
fields of observable supergravity sector, as well as, for dilaton-axion
component of hidden sector of supergravity with their superpartners in
``small"
susy. We will also consider mechanism of spontaneous breaking of susy in the
cosmological models and its influence on the Universe models.
\vfil\eject

\begin{center}
{\LARGE Acknowledgments}
\end{center}

We are grafeful to J. Socorro, M.P. Ryan, O. Obreg\'on and I.C. Lyanzuridi
for their interest in this work. J.J. Rosales is also grateful for support
by Universidad del Bajio, A.C., and CONACyT Graduate Fellowship. This work
was also supported in part by
CONACyT grant 3898P-E9607 and by ISF under grant U96000.

\end{document}